\documentclass{aa}
\usepackage{rotating}
\usepackage{graphicx}
\usepackage{txfonts}
\begin{document}

\title{A multiwavelength investigation of the massive eclipsing binary Cyg\,OB2 \#5\thanks{Based on observations collected at the Observatoire de Haute Provence (France), the Observatorio Astron\'omico Nacional of San Pedro M\'artir (Mexico) and {\it XMM-Newton}, an ESA science mission with instruments and contributions funded by ESA member states and the USA (NASA).}\fnmsep\thanks{Table 1 is only available in electronic form at the CDS via anonymous ftp to cdsarc.u-strasbg.fr (130.79.128.5) or via http://cdsweb.u-strasbg.fr/cgi-bin/gcat?J/A+A/}}
\author{N.\ Linder\inst{1} \and G.\ Rauw\inst{1}\fnmsep\thanks{Research Associate FRS/FNRS (Belgium)} \and J.\ Manfroid\inst{1}\fnmsep\thanks{Research Director FRS/FNRS (Belgium)} \and Y.\ Damerdji\inst{1} \and M.\ De Becker\inst{1}\fnmsep\thanks{Postdoctoral Researcher FRS/FNRS (Belgium)} \and  P.\ Eenens\inst{2} \and \newline P.\ Royer\inst{3} \and J.-M.\ Vreux\inst{1}}
\offprints{G.\ Rauw}
\mail{rauw@astro.ulg.ac.be}
\institute{Institut d'Astrophysique et de G\'eophysique, Universit\'e de Li\`ege, All\'ee du 6 Ao\^ut, B\^at B5c, 4000 Li\`ege, Belgium \and Departamento de Astronom\'{\i}a, Universidad de Guanajuato, Apartado 144, 36000 Guanajuato, GTO, Mexico \and Instituut voor Sterrenkunde, K.U.\ Leuven, Celestijnenlaan 200B, 3001 Leuven, Belgium}
\date{Received date / Accepted date}
\abstract{The properties of the early-type binary \object{Cyg\,OB2 \#5} have been debated for many years and spectroscopic and photometric investigations yielded conflicting results.}{We have attempted to constrain the physical properties of the binary by collecting new optical and X-ray observations.}{The optical light curves obtained with narrow-band continuum and line-bearing filters are analysed and compared. Optical spectra are used to map the location of the He\,{\sc ii} $\lambda$\,4686 and H$\alpha$ line-emission regions in velocity space. New XMM-Newton as well as archive X-ray spectra are analysed to search for variability and constrain the properties of the hot plasma in this system.}{We find that the orbital period of the system slowly changes though we are unable to discriminate between several possible explanations of this trend. The best fit solution of the continuum light curve reveals a contact configuration with the secondary star being significantly brighter and hotter on its leading side facing the primary. The mean temperature of the secondary star turns out to be only slightly lower than that of the primary, whilst the bolometric luminosity ratio is found to be 3.1. The solution of the light curve yields a distance of $925 \pm 25$\,pc much lower than the usually assumed distance of the Cyg\,OB2 association. Whilst we confirm the existence of episodes of higher X-ray fluxes, the data reveal no phase-locked modulation with the 6.6\,day period of the eclipsing binary nor any clear relation between the X-ray flux and the 6.7\,yr radio cycle.}{The bright region of the secondary star is probably heated by energy transfer in a common envelope in this contact binary system as well as by the collision with the primary's wind. The existence of a common photosphere probably also explains the odd mass-luminosity relation of the stars in this system. Most of the X-ray, non-thermal radio, and possibly $\gamma$-ray emission of Cyg\,OB2 \#5 is likely to arise from the interaction of the combined wind of the eclipsing binary with at least one additional star of this multiple system.}
\keywords{Stars: early-type -- Stars: individual: Cyg\,OB2 \#5 -- Stars: binaries: eclipsing -- Stars: winds, outflows -- Stars: fundamental parameters -- X-rays: stars}
\maketitle

\section{Introduction \label{introduction}}
The eclipsing nature of Cyg\,OB2 \#5 (= V\,729\,Cyg = BD\,$+40^{\circ}$\,4220) was first reported by Miczaika (\cite{Miczaika}). Photometric data were subsequently published by Hall (\cite{Hall}), who gathered 143 $UBV$ observations. Hall (\cite{Hall}) noted the existence of intrinsic variability as well as a slight asymmetry in the light curve which made the system appear brighter during the maximum that followed the secondary eclipse. The light curve obtained by Hall (\cite{Hall}) revealed unequal eclipse depths, thus suggesting that the primary should have a larger surface brightness than the secondary. 

From the analysis of a set of optical spectra, Bohannan \& Conti (\cite{BC}) concluded that the two components of Cyg\,OB2 \#5 should be of equal brightness, whilst they simultaneously inferred a mass ratio (primary/secondary) of $q = 4.3$. Therefore, the secondary star would appear to be far too luminous for its mass, and these authors suggested accordingly that the secondary might be an O-star on its way to becoming a Wolf-Rayet object. A re-analysis of the same spectra by Massey \& Conti (\cite{MC}) yielded a somewhat lower mass ratio of 3.3, but did not alter the basic conclusions of the Bohannan \& Conti (\cite{BC}) paper.

Leung \& Schneider (\cite{LS}) analysed Hall's (\cite{Hall}) $UBV$ light curve adopting the mass ratio proposed by Bohannan \& Conti (\cite{BC}). Their description of the system was that of an overcontact binary with the primary accounting for almost 90\,\% of the total light, unlike the almost equal luminosities inferred from the strengths of the He absorption lines (Bohannan \& Conti \cite{BC}). Whilst the luminosity difference of 2.1\,mag inferred by Leung \& Schneider (\cite{LS}) is in qualitative agreement with a large mass ratio, it is clearly at odds with the strong spectral signature of the secondary in the spectrum of the binary. 
 
Vreux (\cite{JMV}) studied the variations in the H$\alpha$ emission line profile and suggested that the primary is currently losing material towards the secondary. This was interpreted as further evidence that the secondary star could be a Wolf-Rayet type object with an unusual spectrum produced by the accretion of the material. 

Rauw et al.\ (\cite{Rauw}) analysed an extensive set of optical spectra and presented a revised orbital solution along with a discussion about the phase-locked spectroscopic variability. The new orbital solution yielded a mass ratio of $q = 3.31 \pm 0.25$ and revealed a large difference in the systemic velocities of the primary and the secondary (the latter having a more negative $\gamma$). Based on the relative strengths of classification absorption lines, Rauw et al.\ (\cite{Rauw}) assigned an O6.5-7 spectral type to the primary and Ofpe/WN9 to the secondary. They also compared the strengths of the primary and secondary absorption lines to infer a visual brightness ratio of $1.4 \pm 0.6$. The complex behaviour of the absorption lines (some of which go into emission during part of the orbit) prevents however a more accurate determination of this ratio. The radial velocities of the emission lines imply that the emissions most likely originate in material that is moving from the primary towards the secondary. Rauw et al.\ (\cite{Rauw}) also reported on variations in the equivalent width of the He\,{\sc ii} $\lambda$\,4686 emission line, which suggest an occultation effect of the emission zone in addition to the modulation of the continuum light curve due to the photometric eclipses. The variations in the line EW suggest that the emission arises from a region close to the secondary star, slightly preceding this star on its orbit. One of the main motivations of the present photometric campaign was to study this behaviour using a set of specially designed narrow-band photometric filters. 

Another goal of this paper is to study the variability of the X-ray emission of Cyg\,OB2 \#5. X-ray emission from this star was first discovered with the {\it EINSTEIN} satellite in December 1978 during a calibration observation of Cyg\,X-3 (Harnden et al.\ \cite{Harnden}). Since then, our target has been observed with many X-ray satellites, including {\it ROSAT} (Waldron et al.\ \cite{Waldron}) and {\it ASCA} (Kitamoto \& Mukai \cite{KM}). We conducted an observing campaign with {\it XMM-Newton} to improve our knowledge of the X-ray emission of Cyg\,OB2 \#5.

Finally, we emphasize that important information about this enigmatic system can be derived from the radio domain. Indeed, Miralles et al.\ (\cite{Miralles}) noted that the radio emission of Cyg\,OB2 \#5 alternates between a low state with a spectral index that is roughly consistent with thermal wind emission and a high state with a much flatter (partially non-thermal) spectral index on a $\sim 7$\,yr period. Abbott et al.\ (\cite{Abbott}) and Miralles et al.\ (\cite{Miralles}) reported on the existence of a radio companion 0.8\,arcsec to the NE of the main radio source (which is associated with the eclipsing binary). Subsequent observations by Contreras et al.\ (\cite{Contreras}) revealed that this radio companion has an elongated shape and lies in-between the short-period binary and a third star\footnote{The latter star is 3 -- 4\,mag fainter than the binary system, and its colours and magnitude suggest a B0-B2\,V spectral type. The angular separation between this astrometric companion and the close binary system suggests an orbital period of several thousand years.}, which was first reported by Herbig (\cite{Herbig}). Contreras et al.\ (\cite{Contreras}) suggested accordingly that the radio companion corresponds to the wind interaction zone between the binary system and the tertiary component. Recently, Kennedy et al.\ (\cite{Kennedy}) re-analysed all VLA observations of Cyg\,OB2 \#5 and showed that the primary radio source, associated with the eclipsing binary varies on a period of $6.7 \pm 0.2$\,yr whilst the secondary source remains constant. The variations in the main radio component can be represented by a simple model in which a fourth (unresolved) object orbits the eclipsing binary in an eccentric orbit and the varying radio emission results from a variable non-thermal emission produced in the wind interaction between the short period binary and the fourth star (Kennedy et al.\ \cite{Kennedy}).
 
\section{Observations \label{observations}}
\subsection{Photometry}
Our photometric monitoring campaign was conducted over 14 nights in August - September 1998 (between HJD\,2451047.3 and 2451060.6) at the OHP 1.20\,m telescope. The telescope was equipped with a $1024 \times 1024$ SITe CCD camera. The pixel size was $24$\,$\mu$m corresponding to 0.69\,\arcsec on the sky. We used a set of five filters with narrow bandpasses designed especially for the photometric search and classification of Wolf-Rayet stars (see Royer et al.\ \cite{Pierre}). Three of these filters are centred on lines that are in emission in various subtypes of Wolf-Rayet objects (He\,{\sc ii} $\lambda$\,4686, He\,{\sc i} $\lambda$\,5876, and C\,{\sc iv} $\lambda\lambda$\,5801, 5812), whilst the other two (hereafter $c_1$ and $c_2$ respectively) are centred on line-free regions around 5057\,\AA\ and 6051\,\AA\ (see Royer et al.\ \cite{Pierre} and their Table\,1 for further details). The exposure times varied between 1 and 2 minutes depending on the atmospheric conditions.

The data were reduced independently by two of us (JM and YD) with different software tools. The results of both reductions were found to be in good agreement. The individual images were bias subtracted, divided by a median (calculated for each night and each filter), normalized flat-field and recentered in a homogeneous way. We used the {\sc daophot} software (Stetson \cite{Stetson}) with aperture radii of 5, 7, 10, and 16 pixels (corresponding to 3.5, 4.8, 6.9, and 11 arcsec).  Whilst the results for the different apertures are generally in good agreement, we focus here on the light curves obtained for a 7-pixel radius aperture, which is well adapted to the general seeing conditions at the Haute-Provence site. The astrometric positions of the sources were obtained from combinations of the highest quality individual frames. Absolute (all-sky) photometry was derived from the larger aperture data, using a multi-night, multi-filter algorithm and a few standard stars (Manfroid \cite{JM}). This procedure yielded additional reference stars for each field. These secondary standards and all non-variable stars were used to determine by a global minimization procedure, the zeropoints for the individual frames and each aperture radius, thus performing global differential photometry. Compared with the best-fit of the light curve (see below), we found a strange systematic trend in the O - C residuals of all filters, with the star progressively becoming fainter (by about 0.02\,mag) towards the end of the night (i.e.\ at higher airmasses). We carefully checked all reduction steps and found no reason for this behaviour. No comparison star was found to exhibit the same behaviour. In a Fourier periodogram of the residuals, we detected a peak at 1.01\,d$^{-1}$. Whilst we were unable to identify the origin of this effect, which could thus be real, we note that (perhaps by coincidence) the effect disappeared when we discarded data taken at airmasses higher than 1.3. This is therefore the strategy that we adopt throughout this paper.

\subsection{Spectroscopy}
In September 1996 and April 1997, we obtained eight spectra of Cyg\,OB2 \#5 with the echelle spectrograph mounted on the 2.1\,m telescope at the Observatorio Astron\'omico Nacional of San Pedro M\'artir (SPM) in Mexico. The spectrograph covers the spectral domain between about 3800 and 6800\,\AA\ with a resolving power of 18\,000 at 5000\,\AA. The detector was a SITe CCD with $1024 \times 1024$ pixels of 24\,$\mu$m$^2$. The data were reduced using the echelle context of the {\sc midas} software, and specific orders covering important lines (H$\beta$, H$\alpha$, He\,{\sc i} $\lambda$\,5876, He\,{\sc ii} $\lambda$\,4686, C\,{\sc iii} $\lambda$\,5696) were normalized using carefully chosen continuum windows. 

\subsection{X-ray observations}
We obtained six {\it XMM-Newton} observations centred on Cyg\,OB2 \#8a, including Cyg\,OB2 \#5 in the field of view of the EPIC cameras (Turner et al.\ \cite{Turner}, Str\"uder et al.\ \cite{Strueder}). The first four pointings separated by 10 days each were obtained in October - November 2004. Two follow-up observations were gathered in April - May 2007. The raw data were processed with the EPIC pipeline chains of the Science Analysis System (SAS) software version 6.0. Some bad time intervals characterised by high background events (so-called soft-proton flares) were rejected. For a detailed description of the data reduction procedure, we refer to De Becker et al.\ (\cite{Michael}). A few stray-light features (due to singly reflected photons) from Cyg\,X-3 are visible in the EPIC field of view. However, they do not affect the position of our target. We extracted the EPIC spectra of Cyg\,OB2 \#5 whenever the source was unaffected by CCD gaps or bad columns. All the spectra were subsequently analysed using the {\sc xspec} software (Arnaud \cite{Arnaud}).

We also retrieved archival {\it ROSAT}-PSPC and {\it ASCA}-SIS observations. Two PSPC data sets are available, rp200109n00 and rp900314n00. The latter was spread over a bit less than five days and we thus divided it into four time bins to search for orbital variability in the X-ray flux of Cyg\,OB2 \#5. For {\it ASCA}, a single observation, obtained during the performance verification phase was obtained. This data set was already discussed by Kitamoto \& Mukai (\cite{KM}). All the archive data were reduced using the {\sc xselect} package. Spectra of the source and the background were extracted, the latter over a source free region close to our target. The response matrices and ancillary response files were either retrieved from the archive or generated by the appropriate {\sc ftools} routines. 

\section{Results \label{results}}
\subsection{The optical light curve}
\subsubsection{The ephemeris of Cyg\,OB2 \#5}
In the literature, the ephemeris of the primary eclipse was given by $HJD = 2440413.796 + 6.5977915\,E$ (Hall \cite{Hall}). H\"ubscher \& Walter (\cite{HW}) reported the time of the minimum of the light curve as HJD\,$2453985.4928 \pm 0.0015$ based on 15 measurements obtained in 2006. The time of minimum published by H\"ubscher \& Walter (\cite{HW}) is 2057.006 cycles after the zero time of Hall (\cite{Hall}). From our data, we infer HJD\,$2451049.702 \pm 0.025$ to be the time of primary minimum, which corresponds to a shift of 0.04 in phase compared with Hall's ephemeris (1612 cycles). This apparent discrepancy prompted us to re-examine the question of the system's orbital period. Using the time of conjunction from the radial velocity curve of Wilson \& Abt (\cite{WA}) and the times of photometric minima listed by Miczaika (\cite{Miczaika}), Hall (\cite{Hall}), H\"ubscher \& Walter (\cite{HW}) as well as our own data, we found that the best fit linear ephemeris was $HJD = (2440413.750 \pm 0.011) + (6.597840 \pm 0.000009)\,E$. However, these new ephemerides lead to uncomfortably large phase shifts when applied to the epochs of our observations. Therefore, since the H\"ubscher \& Walter (\cite{HW}) result is apparently based on only 15 data points, we discarded this point from the fit\footnote{Given the difficulties in defining accurately the time of primary minimum in our data (containing several hundred measurements), the error in the time of minimum quoted by H\"ubscher \& Walter (\cite{HW}) seems extremely optimistic to us.}. A revised linear ephemeris would then become $HJD = (2440413.718 \pm 0.013) + (6.597888 \pm 0.000011)\,E$. At first sight these new ephemerides might look less precise than the old ones, but we emphasize that Hall (\cite{Hall}) did not provide an error bar in his ephemeris. Since the system is likely to be in an evolutionary phase with complex mass exchanges and mass loss (see below), we also considered the possibility of quadratic ephemerides. Using the times of photometric minima apart from the H\"ubscher \& Walter (\cite{HW}) value, we obtain 
\begin{eqnarray*} 
HJD & = & HJD_0 + P_0\,E + \frac{1}{2}\,\dot{P}\,P_0\,E^2 \\
& = & (2440413.796 \pm 0.024) + (6.597858 \pm 0.000014)\,E \\
& & + (6.11 \pm 1.64) \times 10^{-8}\,E^2
\end{eqnarray*}
These are the ephemerides that we use throughout this paper. The orbital period at the time of our photometric observations would hence be $P = 6.598055$\,days, whilst the time derivative of the period would amount to $\dot{P} = (18.5 \pm 5.0) \times 10^{-9}$\,s\,s$^{-1} = (0.58 \pm 0.16)$\,s\,yr$^{-1}$.
\subsubsection{The continuum filters}
We first studied the light curve of Cyg\,OB2 \#5 in the $c_1$ and $c_2$ continuum filters\footnote{Table 1 with the light curves of Cyg\,OB2 \#5 in the various filters is available in electronic form at the CDS.}. Restricting ourselves to observations taken at an airmass below 1.3, the total number of usable data points was 320 and 299 for the $c_1$ and $c_2$ filters, respectively. The data provided a good coverage of the orbital cycle, although they did not sample the core phase of secondary minimum. These light curves indeed confirm that both minima have different depths and also indicate a different brightness immediately after the primary eclipse compared to the situation after the secondary eclipse. These features agree with the description of Hall (\cite{Hall}) and cannot be explained by stars with a uniform surface brightness. We thus considered situations where at least one of the two stars has a non-uniform brightness distribution (see below). To analyse this light curve further, we used the {\sc nightfall} software\footnote{ http://www.hs.uni-hamburg.de/DE/Ins/per/Wichmann/\\Nightfall.html} developed and maintained by R.\ Wichmann, M.\ Kuster and P.\ Risse. 

Rauw et al.\ (\cite{Rauw}) derived two slightly different values of the dynamic mass ratio $q = m_p/m_s = 3.31 \pm 0.25$ when only the new RV data were used in the orbital solution and $q = 3.54 \pm 0.18$ when literature RV data were also included. In a first series of trials, we hence tested the capabilities of the light curve analysis to distinguish between these different values of the mass ratio. We found that within a $3\,\sigma$ interval around the mean mass ratios, the light curve fits were not very sensitive to the value of $q$. However, the fits for the lower value of $q = 3.31$ are of somewhat better quality and in the following we thus keep the mass ratio fixed at the value derived from the new RV data of Rauw et al.\ (\cite{Rauw}). 

At first, we tested models with the surface temperature distribution ruled by gravity darkening only (i.e.\ no bright or dark spots). The {\sc nightfall} code immediately converges towards a contact (or even over-contact) configuration where both stars fill up their Roche lobe and are seen under an inclination $i$ close to $62$ -- $67^{\circ}$. The best-fit model temperature of the secondary star is 29\,500\,K in this case (assuming a primary effective temperature of 36\,000\,K).

The residuals of this solution indicate that the data points during the egress of the primary eclipse are systematically lower than the model predictions, whilst the agreement between the data and the model is reasonable over the remainder of the light curve. We note that allowing for a slightly eccentric orbit does not solve this discrepancy between the data and the synthetic light curve. This feature could be explained instead by either the secondary being somewhat brighter on its leading side or the primary being fainter on its leading side. Since the first option is reminiscent of the location of the colliding wind region discussed by Rauw et al.\ (\cite{Rauw}), we tested this model by adding a bright ``spot'' on the secondary\footnote{We note that we assumed the spot to be on the stellar equator, which should be a reasonable assumption since the hot region is likely to arise from mutual heating effects and/or interactions with the primary's wind.}. This bright spot model yields a significantly better fit. However, there is a degeneracy between the secondary surface temperature, spot size, and spot temperature ratio. These parameters combine in such a way as to produce a mean temperature of the secondary star over its entire surface of about 35\,600\,K. As a result, the secondary's surface temperature outside the spot area is no longer constrained and we thus decided to adopt the value of 29\,500\,K derived above. We then find that the part of the secondary star facing the primary is significantly hotter (the hot spot has a best-fit model temperature of 36\,500\,K, as compared to 29\,500\,K for the rest of the secondary's surface) and hence brighter (by a factor of 2.35 in bolometric luminosity). Actually, the surface of the ``spot'' covers more than half of the total surface of the secondary star! However, the centre of this hot spot is found not to be aligned with the binary axis. In fact, the best-fit model yields a longitude for the spot centre of $-15^{\circ}$, where $0^{\circ}$ corresponds to the binary axis and negative longitudes refer to the leading side of the secondary's surface. The fact that the centre of the spot is located on the leading side of the secondary star suggests indeed that it cannot be due to a sole heating effect by the primary's radiation, but that additional heating from a Coriolis-deflected wind interaction region wrapped around the secondary star is probably playing a major role. This situation is in excellent qualitative agreement with the expectations for the wind interaction zone characterised by variations in the equivalent width of the He\,{\sc ii} $\lambda$\,4686 emission (Rauw et al.\ \cite{Rauw}).

\addtocounter{table}{1}
\begin{table}[htb]
\begin{minipage}[t]{\columnwidth}
\begin{center}
\caption{Best-fit parameters of the continuum light curves of Cyg\,OB2\#5. \label{Bestfit}} \renewcommand{\footnoterule}{}  
\begin{tabular}{l c}
\hline
Parameter & $c_1$ \& $c_2$ \\
\hline
$i$ ($^{\circ}$) & $64.2^{+1.3}_{-1.7}$ \\
$q$              & $3.31$ (fixed) \\
Filling factor primary & $1.00^{+.02}_{-.01}$ \\
Filling factor secondary & $1.00^{+.02}_{-.01}$ \\
$T_{\rm eff, p}$\,(K) & 36000 (fixed)\\
$T_{\rm eff, s}$\,(K) & 29500 (fixed\footnote{This value was derived from the best-fit model without a spot on the secondary star and refers here only to the un-spotted part of the secondary's surface.}) \\
\hline
Bright ``spot'' & \\
Longitude & $(-15^{+12}_{-16})^{\circ}$\\
Latitude  & $0^{\circ}$ (fixed) \\
Radius    & $(130^{+11}_{-29})^{\circ}$ \\
Temperature ratio & $1.24^{+.21}_{-.14}$ \\
\hline
\end{tabular}
\end{center}
\end{minipage}
\end{table}

The mean values of the absolute differences between the observed and computed magnitudes $|O - C|$ are $0.009$\,mag for both filters. These values are slightly, but significantly larger than the standard deviations (rms) of the differential magnitudes of non-variable stars of comparable magnitude, probably revealing intrinsic variability of the system in addition to the systematic orbital effects.
\begin{figure}[htb!]
\resizebox{9.0cm}{!}{\includegraphics{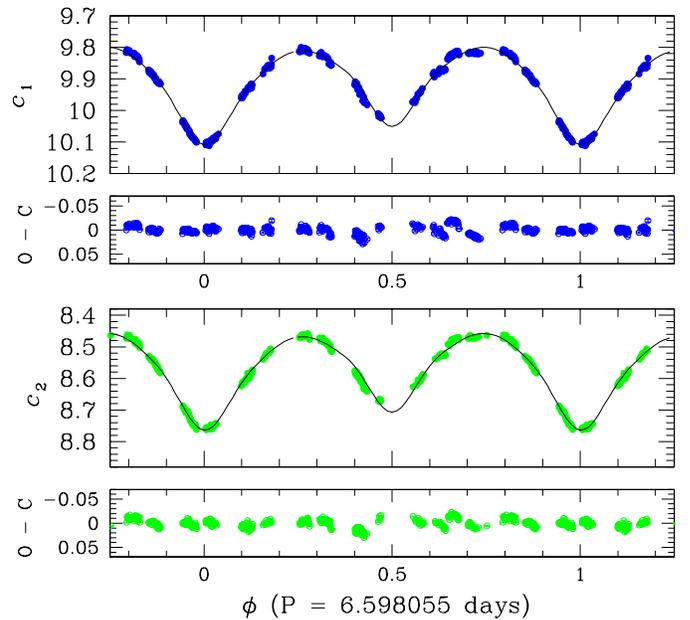}}
\caption{The $c_1$ and $c_2$ light curves of Cyg\,OB2 \#5 along with our best-fit model (Table\,\ref{Bestfit}). The phases were computed using our quadratic ephemeris, and phase $\phi = 0.0$ corresponds to the primary minimum (secondary star in front). The panels below the light curves illustrate the deviations $O - C$ for each filter.\label{lcc1c2}}
\end{figure}

To establish the error bars on the best-fit parameters in Table\,\ref{Bestfit}, we explored the parameter space by keeping one of the parameters fixed and allowing the others to vary until a minimum in $\chi^2$ was reached. The error bars were then calculated for the variation in $\chi^2$ corresponding to a 90\% confidence level and the appropriate number of degrees of freedom. We emphasize that the parameter space is complex and some parameters (such as the spot's temperature ratio and its size) are correlated to some extent. 
\begin{figure}[htb!]
\resizebox{9.0cm}{!}{\includegraphics{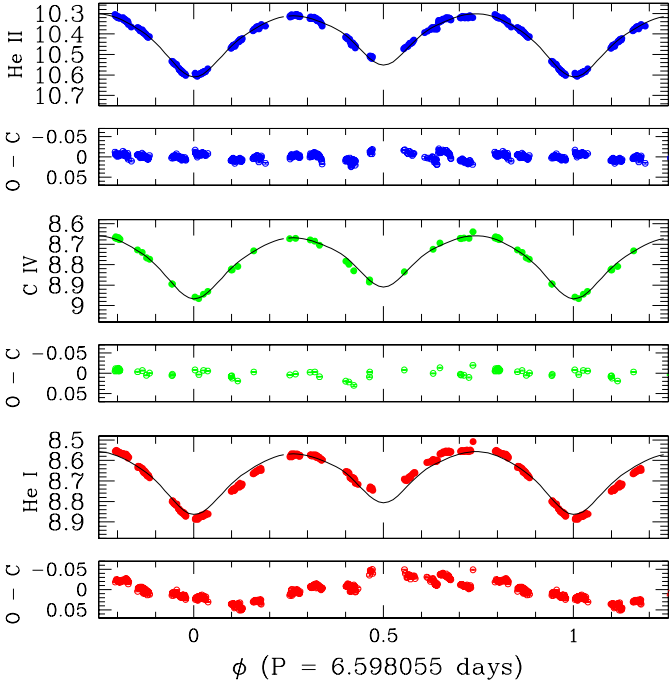}}
\caption{The light curves of Cyg\,OB2 \#5 as observed through the line-bearing filters. The solid line yields the best fit solution of the $c_1$ and $c_2$ filters (Table\,\ref{Bestfit}). The panels below the light curves illustrate the deviations between the actual data and the best fit solution for the continuum filters.\label{lclines}}
\end{figure}

From the orbital solution of Rauw et al.\ (\cite{Rauw}) and the inclination of $64.2^{\circ}$, we infer the absolute masses of the components of Cyg\,OB2 \#5 to be $(31.9 \pm 3.2)$\,M$_{\odot}$ for the primary and $(9.6 \pm 1.1)$\,M$_{\odot}$ for the secondary. With mean (averaged over the entire surface) surface temperatures of 36\,000 and 35\,600\,K for the primary and secondary respectively, the bolometric luminosity ratio between the two stars is about 3.1, whilst the optical ($V$-band) brightness ratio should be around 2.9. The total bolometric luminosity of the system equals about $\log{L_{\rm bol}/L_{\odot}} = 6.04$ corresponding to an absolute magnitude of $M_V = -7.05$. The observed $V$ magnitude outside eclipse is $9.20$ around maximum (Hall \cite{Hall}), whilst it is 9.80 for the $c_1$ filter\footnote{Using the Cardelli et al.\ (\cite{Cardelli}) reddening law, we infer $A_{c1}/A_V = 1.106$.}. For the reddening of Cyg\,OB2 \#5, Torres-Dodgen et al.\ (\cite{Torres}) quoted $A_V = 6.40$, whilst Wegner (\cite{Wegner}) gave a value of $A_V = 6.35$. These numbers also agree with the average $B - V = 1.68$ value of the binary inferred from the data compiled by Reed (\cite{Reed}). Hence, our analysis of the light curve yields a distance modulus in the range 9.77 -- 9.90 corresponding to a distance of about 900 -- 950\,pc. This is significantly less than the value of Torres-Dodgen et al.\ (\cite{Torres}), who used $ubvyJHKL'$ photometry of the Cyg\,OB2 association to infer a distance modulus of $11.2 \pm 0.2$. The latter value is in good agreement with the distance modulus of 11.2 proposed by Massey \& Thompon (\cite{MT}). It is obvious that our result depends strongly on the (assumed) temperature of the primary and secondary components. Here, we adopted 36\,000\,K for the primary as inferred by Rauw et al.\ (\cite{Rauw}) from the O6.5-7 spectral type. However, to ensure that our results agree with the 1.7\,kpc distance estimate, we would require a much hotter primary (of order 48\,000\,K) which appears extremely unlikely. On the other hand, our distance estimate is not very sensitive to uncertainties in the secondary's temperature. 

Whilst our best-fit parameters differ significantly from those of Leung \& Schneider (\cite{LS}), we emphasize that a similar discrepancy between the luminosity of the light curve fit and that inferred from the typical distance was also noted by the latter authors, although they formulated this problem in a different way. We therefore encourage other studies of eclipsing binaries in Cyg\,OB2 (e.g.\ Cyg\,OB2 \#3; see Kiminki et al.\ \cite{Kiminki}) to confirm independently the distance of the cluster. 

\subsubsection{Line-bearing filters}
We then compared the best-fit solutions of the continuum filters with the observed light curves of the filters centred on the lines. The results are shown in Fig.\,\ref{lclines}. For each of the three line-bearing filters, we simply shifted the theoretical $c_1$ or $c_2$ light curve until it fitted the average magnitude measured in the line-bearing filter. The remaining residuals then inform us about effects that are specific to a given filter. No strong effect is evident for filters centred on He\,{\sc ii} $\lambda$\,4686 (306 data points, $\overline{|O - C|} = 0.008$\,mag) and C\,{\sc iv} $\lambda\lambda$\,5801, 5812 (44 data points, $\overline{|O - C|} = 0.010$\,mag). The average residuals are fully consistent with the results obtained for the $c_1$ and $c_2$ filters. In fact, the spectrum of Cyg\,OB2 \#5 does not contain any strong emission or absorption lines in the C\,{\sc iv} $\lambda\lambda$\,5801, 5812 filter. Hence, it is unsurprising that this filter behaves in the same way as the continuum filters. However, a very different situation is observed for the filter centred on He\,{\sc i} $\lambda$\,5876, for which we obtain $\overline{|O - C|} = 0.022$\,mag (312 data points) and Fig.\,\ref{lclines} clearly reveals a phase dependent modulation of the residuals. Actually, the light deficit after primary minimum appears stronger in this filter than in any other filter that we have used. As a result, the residuals indicate a broad minimum around phase 0.15. We therefore attempted to fit the light curves of the three line-bearing filters with the same model as for the continuum filters, but allowing the secondary's temperature as well as the spot's temperature ratio, longitude, and size to vary until we achieve the best-fit for each filter. The results are listed in Table\,\ref{linesfit}.

\begin{table}[htb]
\begin{minipage}[t]{\columnwidth}
\begin{center}
\caption{Parameters of the best-fit solutions to the light curves of Cyg\,OB2\#5 as observed through the line-bearing filters. \label{linesfit}}
\renewcommand{\footnoterule}{}
\begin{tabular}{l c c c}
\hline
Filter\footnote{Several parameters were frozen in these fits: $i = 64.2^{\circ}$, $q = 3.31$, primary and secondary filling factors = 1.00, $T_{\rm eff, p} = 36000$\,K, latitude of the bright spot $= 0^{\circ}$.} & He\,{\sc ii} & C\,{\sc iv} & He\,{\sc i} \\
\hline
$T_{\rm eff, s}$\,(K) & 28400 & 32600 & 21300 \\
\hline
Bright spot & \\
Longitude $(^{\circ})$ & $-15$  & $-28$  & $-26$  \\
Radius    $(^{\circ})$ & $125$  & $117$  & $118$  \\
Temperature ratio      & $1.14$ & $1.26$ & $1.46$ \\
\hline
\end{tabular}
\end{center}
\end{minipage}
\end{table}

From this table, we find that the apparent bolometric brightness ratio of the spot is essentially the same in the continuum (2.36), He\,{\sc ii} (1.69) and C\,{\sc iv} (2.52) filters, but reaches 4.54 in the He\,{\sc i} filter. We also note that the data for the latter filter yield the lowest secondary temperature, i.e.\ in this filter the primary appears much brighter than the unspotted part of the secondary's surface. We propose an explanation for this behaviour in the next section. 

\begin{figure*}[ht!]
\begin{minipage}{6cm}
\resizebox{6.0cm}{!}{\includegraphics{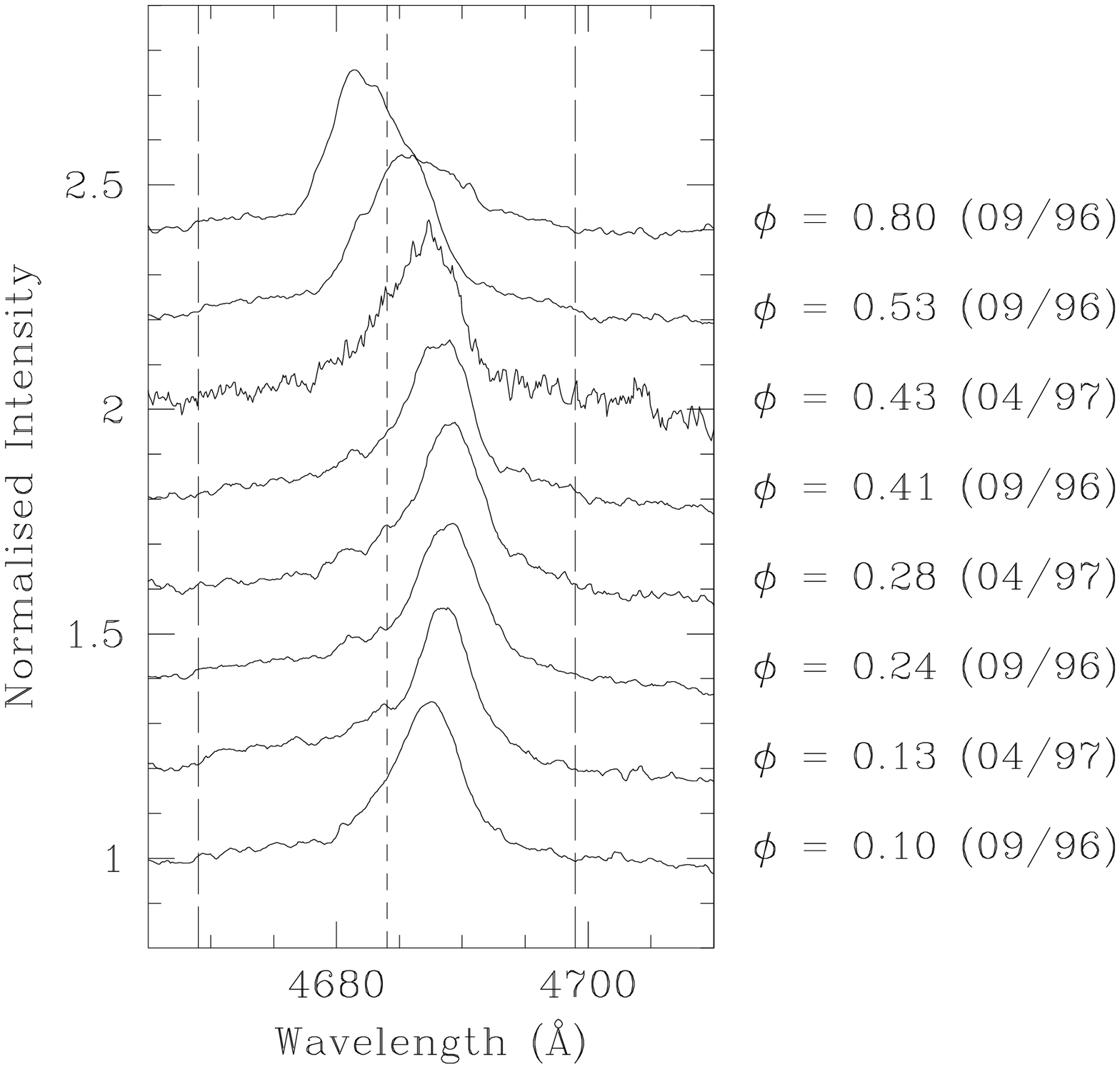}}
\end{minipage}
\hfill
\begin{minipage}{6cm}
\resizebox{6.0cm}{!}{\includegraphics{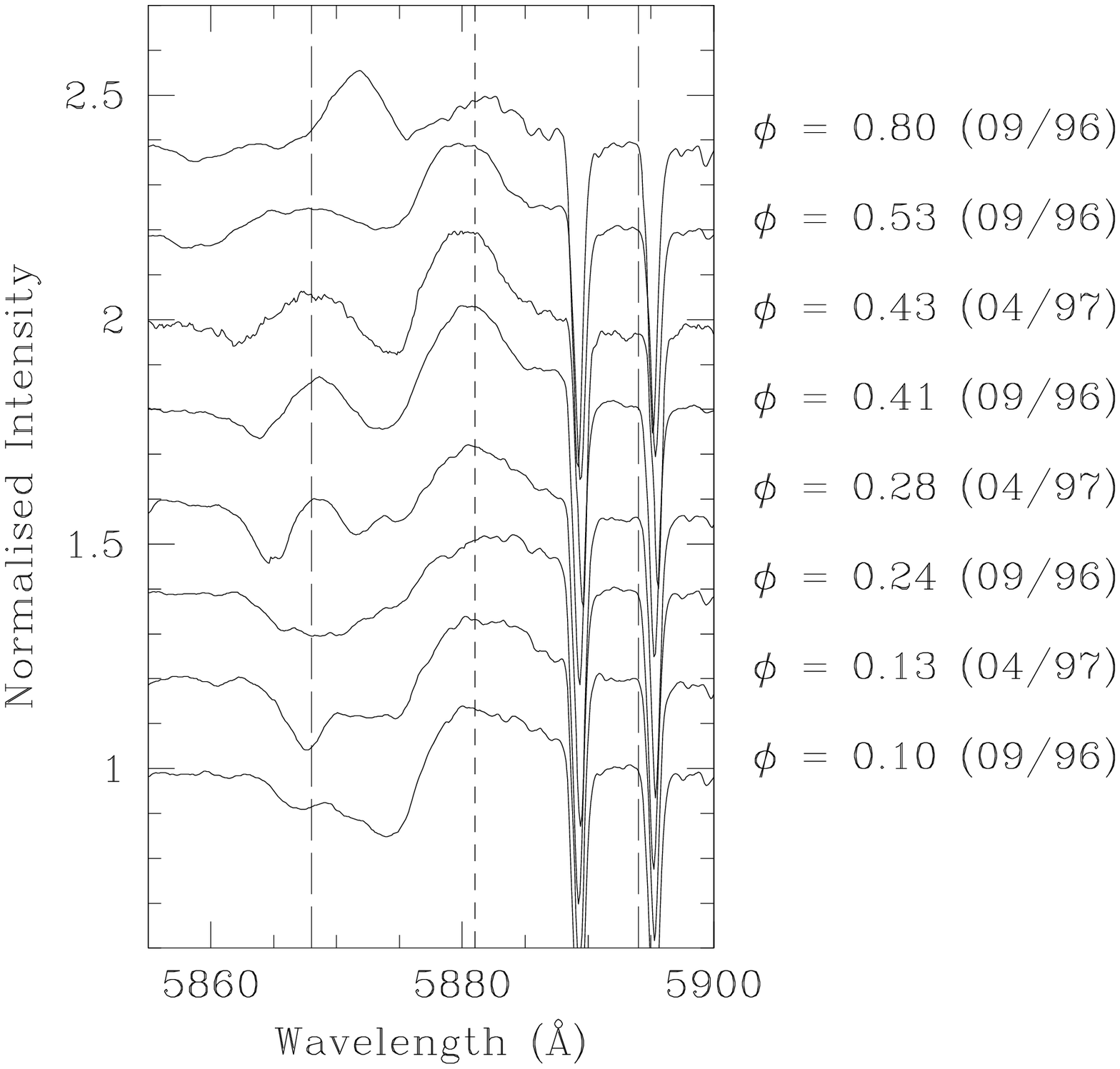}}
\end{minipage}
\hfill
\begin{minipage}{6cm}
\resizebox{6.0cm}{!}{\includegraphics{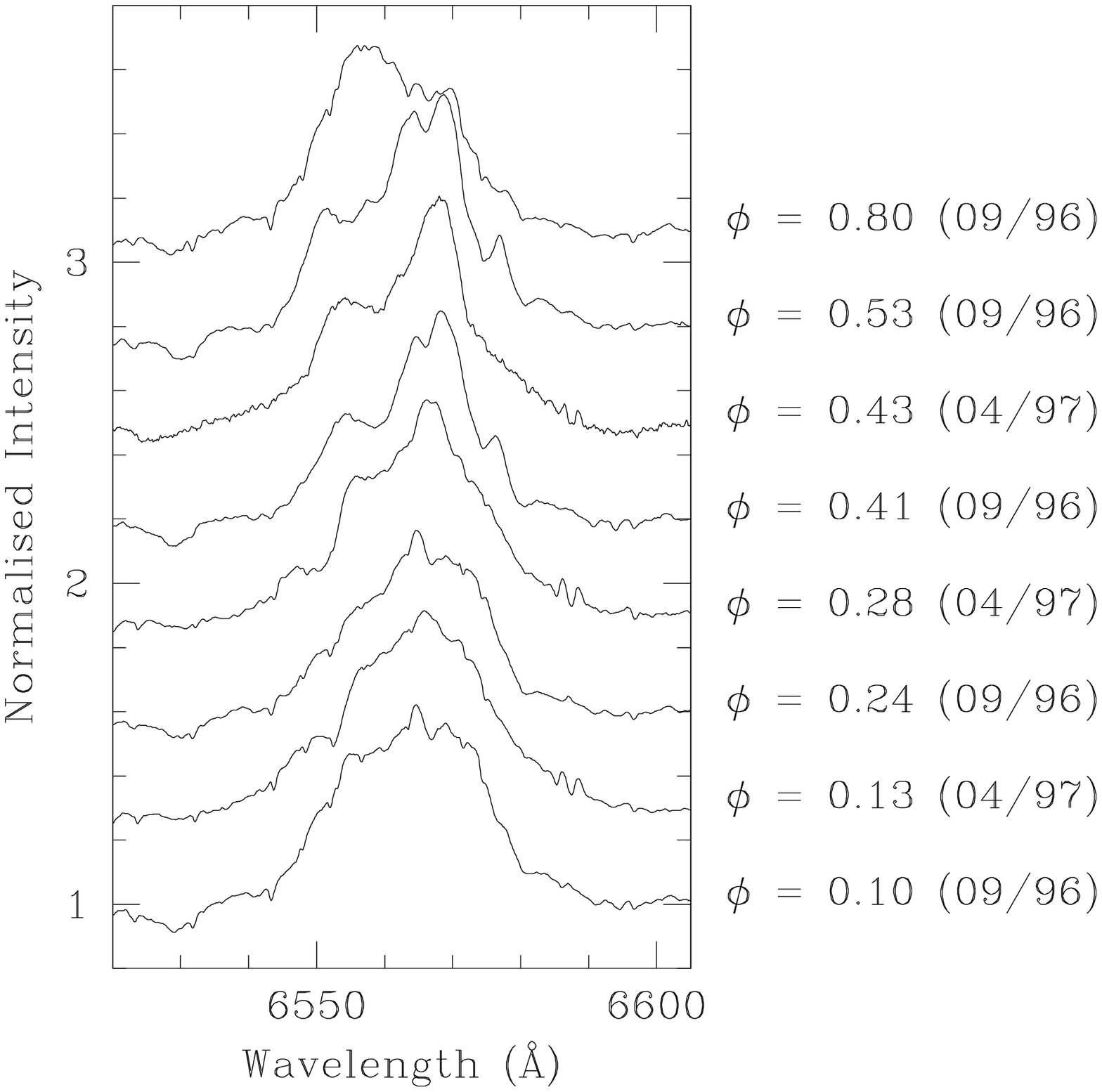}}
\end{minipage}
\caption{Line profile variability in the optical spectrum of Cyg\,OB2 \#5. The panels illustrate from left to right the observed variations in the He\,{\sc ii} $\lambda$\,4686, He\,{\sc i} $\lambda$\,5876, and H$\alpha$ lines. The successive spectra are shifted vertically by 0.2 (respectively 0.3) continuum units in the left and middle (respectively the right) panels. In all cases, the profiles are shown in the heliocentric frame of reference and the intensities of the lines were corrected for the variations in the continuum light curve. In the left and middle panels, the short-dashed lines show the central wavelength of the narrow-band filter centred on these lines, whilst the two long-dashed lines provide the {\tt FWHM} of these filters. The phases of the observations are indicated to the right of the figures (quadratic ephemeris with $\phi = 0.0$ corresponding to the primary minimum). \label{optical}}
\end{figure*}
\subsection{Optical spectroscopy}
Unfortunately, our optical spectra do not provide a complete coverage of the orbital cycle. Nevertheless, we can use them to illustrate the variability of the most prominent lines (see Fig.\,\ref{optical}) and attempt to understand the origin of the discrepant behaviour of the He\,{\sc i} $\lambda$\,5876 filter in our photometric data.

\begin{figure}[h!]
\resizebox{8.0cm}{!}{\includegraphics{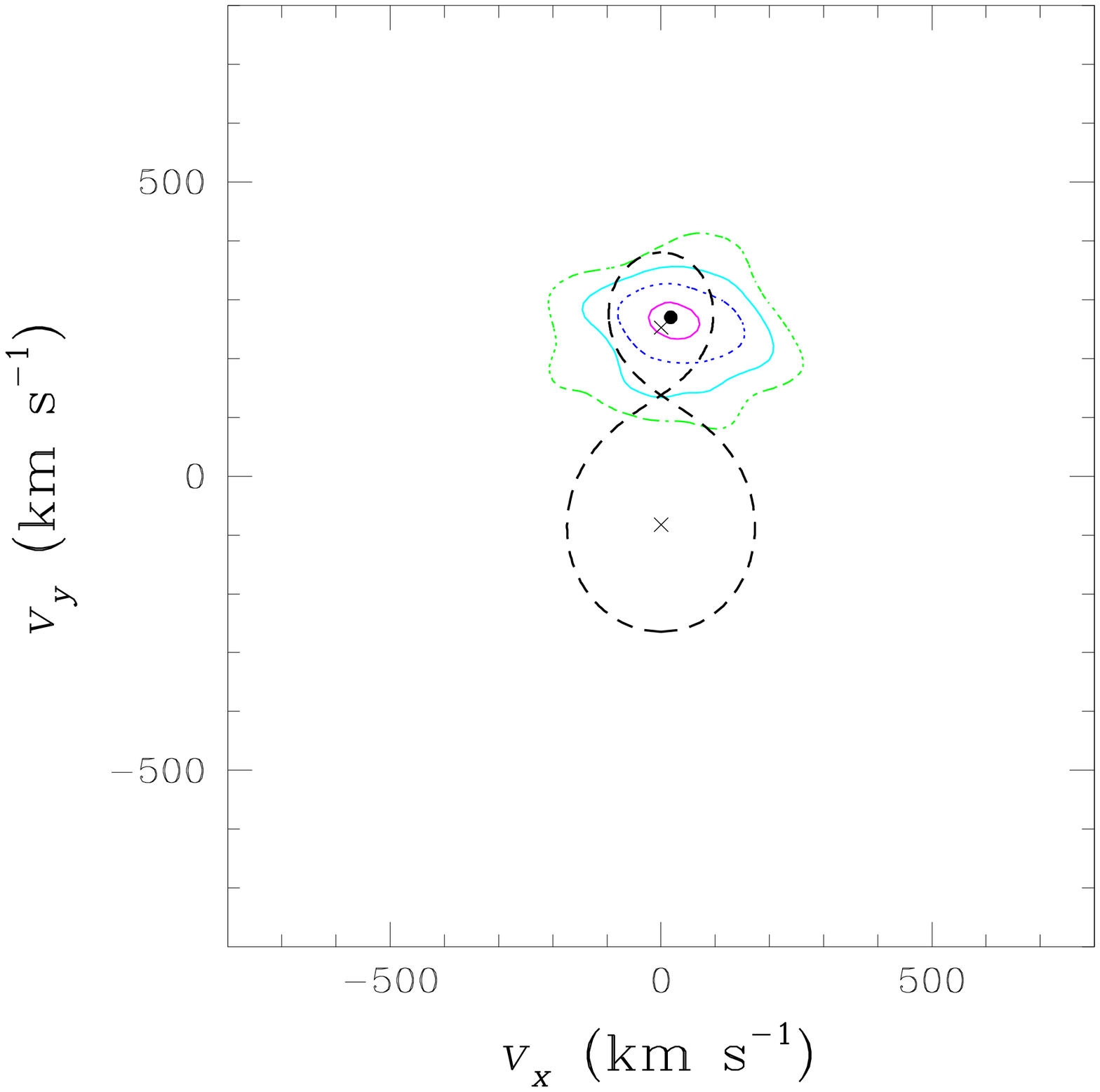}}
\resizebox{8.0cm}{!}{\includegraphics{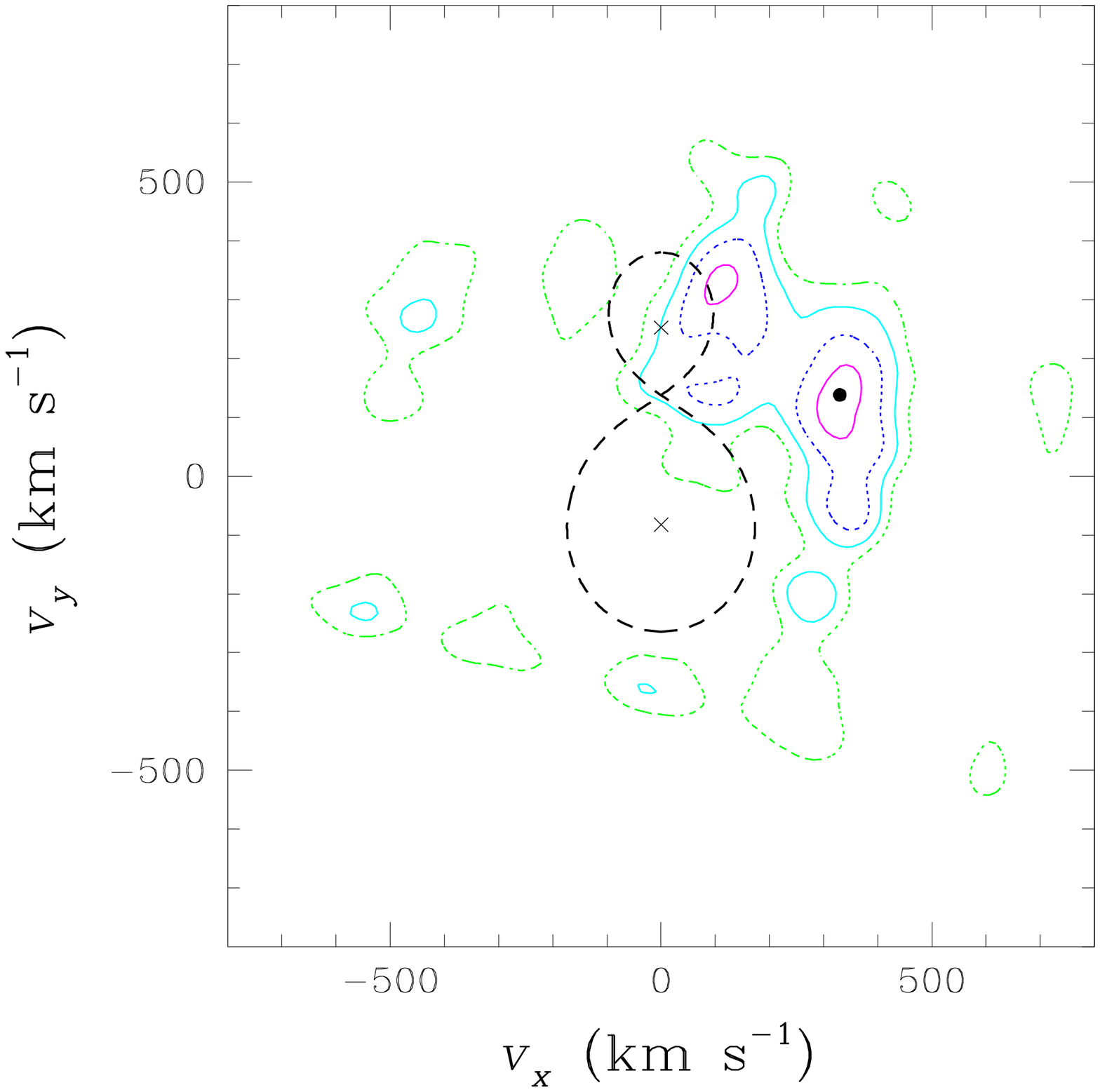}}
\caption{Doppler maps of the He\,{\sc ii} $\lambda$\,4686 (top panel) and H$\alpha$ (bottom panel) lines in the spectrum of Cyg\,OB2 \#5. The crosses correspond to the radial velocity amplitudes of the centre of mass of the primary and secondary. The shape of the Roche lobe in velocity-space (thick dashed line) has been calculated for a mass ratio (primary/secondary) of 3.31. The black dot indicates the position of the highest peak and the contours correspond to levels of 0.95, 0.80, 0.65, and 0.50 times the maximum emissivity. \label{tomo}}
\end{figure}
The variations in the He\,{\sc ii} $\lambda$\,4686 and H$\alpha$ emissions are very similar to the variability pattern reported by Rauw et al.\ (\cite{Rauw}) and  Vreux (\cite{JMV}), respectively. We mapped the emission regions of these emission lines in velocity space using the Doppler tomography technique of Rauw et al.\ (\cite{hde228766}). For this purpose, we adopted a reference frame centred on the centre of mass of the binary with the $x$-axis pointing from the primary to the secondary and the positive $y$-axis pointing along the direction of the secondary's orbital motion. The basic assumption of a Doppler tomography technique is that the phase dependence of the radial velocity $v(\phi)$ of any gas flow that is stationary in the rotating frame of reference of the binary can be described by a so-called `S-wave' relation:
\begin{equation}
v(\phi) = -v_x\,\cos{(2\,\pi\,\phi)} + v_y\,\sin{(2\,\pi\,\phi)} + v_z
\label{eq1}
\end{equation}
where $\phi$ is the orbital phase ($\phi = 0$ at primary eclipse), whilst $v_x$ and $v_y$ are the projected velocity components along the $x$ and $y$ axes and $v_z$ corresponds to the apparent systemic velocity of the line under investigation (assumed to equal the primary's systemic velocity of $-33$\,km\,s$^{-1}$). The data were weighted to account for the uneven sampling of the orbital cycle, and the intensities of the emission lines were corrected for variations in the underlying continuum.

While the limited sampling of our data leads to some low-level artefacts in the Doppler map (suppressed in Fig.\ \ref{tomo} by choosing the lowest contour level at $0.50$ $\times$ the intensity of the highest peak), it can be seen that there is good qualitative agreement between the results obtained here and the simple S-wave analysis of Rauw et al.\ (\cite{Rauw}). For the He\,{\sc ii} $\lambda$\,4686 line, we found a single emission region concentrated close to the velocity of the centre of mass of the secondary star, although with a slightly positive $v_x$. The Doppler map of the H$\alpha$ line is far more complex, reflecting the multi-peak structure of the profile of this line. Nevertheless, the bulk of the emission originates in material with positive $v_x$ and $v_y$ velocities. These emission lines are thus formed mostly in material that moves from the primary towards the secondary and in the same orbital direction as the secondary. Although more complex configurations cannot be excluded at this stage, we note that this picture agrees with a colliding wind scenario where the primary's wind dominates the secondary wind. In such a case, the optical emission lines would arise from recombination in higher density regions due to the impact of the primary wind on the secondary's wind or photosphere. This higher density region would be geometrically coincident with the brighter and hotter region of the secondary star found in our photometric analysis. This description is similar to that presented by Rauw et al.\ (\cite{Rauw}), although we caution that these authors assumed a detached configuration, which is clearly not the case here. The situation in Cyg\,OB2 \#5 is thus quite different from a `classical' wind - wind collision. Indeed, according to our photometric solution, the stars are in contact and the wind interaction mainly concerns those parts of the secondary surface that sweep up the probably more energetic primary wind during the orbital motion. Given the proximity between the components of the binary system, the winds very probably interact at low velocities and the shock-heated material is thus unlikely to reach temperatures sufficiently high to produce a sizeable X-ray emission.

The He\,{\sc i} $\lambda$\,5876 line displays complex variations going from a P-Cygni type profile at phases between 0.1 and 0.4 to an almost pure emission line over the second half of the orbital cycle (see Fig.\,\ref{optical}). This explains the strange behaviour of the photometric data from the filter centred on this line as discussed in the previous section. In fact, the equivalent width of the line, corrected for the photometric variations in the continuum, changes from about 1.0\,\AA\ (for the absorption-dominated P-Cygni type profiles) to $-1.2$\,\AA\ for the emission-dominated profiles. This is quite a significant variation, sufficient to produce the 0.08\,mag (peak-to-peak) deviation of the He\,{\sc i} $\lambda$\,5876 filter from the pure continuum filters. The emission component of the line hence probably arises from a region of limited angular extension, most likely over or above the hot spot on the secondary star. Unfortunately, the highly variable absorption component in this profile prevents us from applying the Doppler tomography technique to the He\,{\sc i} $\lambda$\,5876 line.

In Fig.\,\ref{optical}, we note some hints of slight differences between the profiles observed in 1996 and 1997 at similar orbital phases. In 1997, the absorption trough of the P-Cygni profiles displays a rather sharp absorption component at phase $\phi = 0.13$, whilst no such feature is seen at $\phi = 0.10$ in the 1996 data. This could potentially be related to long-term variations similar to those seen in the H$\beta$ line profile variations (Rauw et al.\ \cite{Rauw}). 

\subsection{X-ray data}
\begin{figure}[htb!]
\resizebox{9.0cm}{!}{\includegraphics{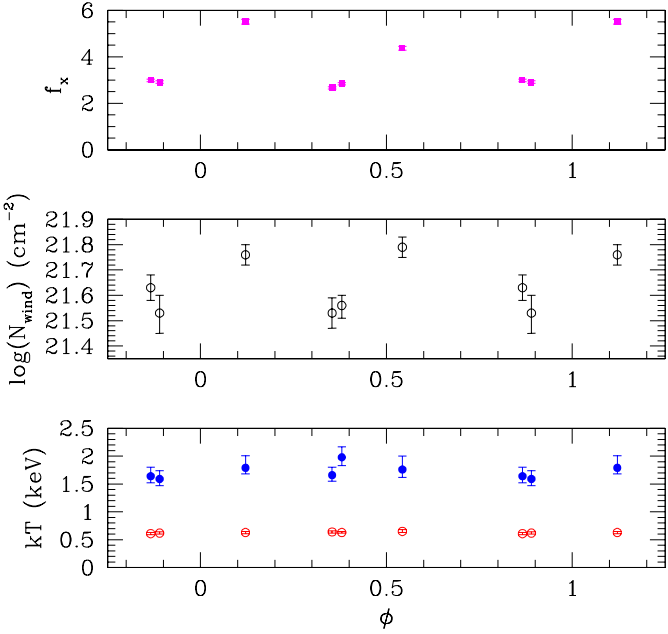}}
\caption{The main parameters of the fits to the EPIC data are shown as a function of phase (quadratic ephemeris with $\phi = 0.0$ corresponding to the primary minimum). From top to bottom, the panels illustrate the observed flux in the 0.4 -- 10.0\,keV band (in units $10^{-12}$\,erg\,cm$^{-2}$\,s$^{-1}$), the circumstellar (wind) column density and the two temperatures ($kT_1$ and $kT_2$) respectively.\label{courbeX}}
\end{figure}

We analysed all the EPIC spectra of Cyg\,OB2 \#5 using the {\sc xspec} software. The spectra were fitted with a two-temperature, optically-thin, thermal plasma model {\tt mekal} (Mewe et al.\ \cite{mewe}, Kaastra \cite{ka}) absorbed by a fixed, neutral interstellar absorption column as well as a column of (partially) ionized circumstellar wind material. Within {\sc xspec}, we hence used a model of the type {\tt wabs$\times$wind$\times$(mekal + mekal)}. The best-fit parameters for the {\it XMM-Newton}-EPIC spectra are listed in Table\,\ref{fitX} and illustrated in Fig.\,\ref{courbeX} for models that assume a solar composition. The first column of Table\,\ref{fitX} yields the dates of the observations (at mid exposure) with the uncertainty corresponding to half the integration time. The interstellar column-density of neutral hydrogen was fixed to the value $1.13 \times 10^{22}$\,cm$^{-2}$ derived from the $E(B - V) = 1.94$ colour excess inferred by Torres-Dodgen et al.\ (\cite{Torres}) and the mean relation between $N_{\rm H}$ and $E(B - V)$ of Bohlin et al.\ (\cite{Bohlin}). The wind component of the absorption was modelled using the opacities computed with the model developed by Naz\'e et al.\ (\cite{HD108}). The norm of the {\tt mekal} model is defined to be $\frac{10^{-14} \times EM}{4\,\pi\,d^2}$ where $d$ is the distance (cm) and $EM = \int n_e\,n_H\,dV$ is the emission measure (in cm$^{-3}$).  The last two columns of Table\,\ref{fitX} yield the observed and ISM absorption corrected fluxes in the 0.4 -- 10.0\,keV energy domain.

\begin{figure}[htb!]
\resizebox{9.0cm}{!}{\includegraphics{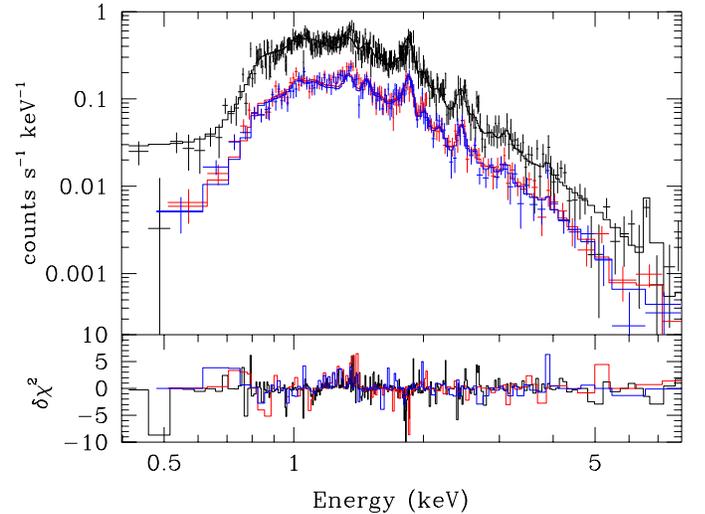}}
\caption{The EPIC-pn (black), -MOS1 (red) and -MOS2 (blue) spectra of Cyg\,OB2 \#5 as observed on JD\,2453328.543 (18 November 2004). The data points are shown with their error bars, whilst the continuous line yields the best-fit model (see Table\,\ref{fitX}). The bottom panel yields the contributions of the individual energy bins to the $\chi^2$ of the fit with the sign of the corresponding $O - C$ difference.\label{fit}}
\end{figure}

\begin{table*}[htb]
\caption{Results of the best-fit models of the EPIC spectra of Cyg\,OB2 \#5. 
\label{fitX}}
\begin{center}
\begin{tabular}{c c c c c c c c c}
\hline
JD - 2400000 & $\log{{\rm N}_{\rm wind}}$ & kT$_1$ & norm$_1$ & kT$_2$ & norm$_2$ & $\chi^2_{\nu}$ (d.o.f.) & f$_X^{\rm obs}$ & f$_X^{\rm corr}$ \\
& & keV & & keV & & & erg\,cm$^{-2}$\,s$^{-1}$ & erg\,cm$^{-2}$\,s$^{-1}$ \\
\hline
\vspace*{-3mm}\\
$53308.579 \pm 0.122$ & $21.53^{+.06}_{-.06}$ & $0.64^{+.02}_{-.02}$ & $(9.04^{+1.09}_{-0.95}) \times 10^{-3}$ & $1.66^{+.14}_{-.11}$ & $(2.62^{+.32}_{-.34}) \times 10^{-3}$ & 1.40 (472) & $2.66 \times 10^{-12}$ & $1.46 \times 10^{-11}$ \\
\vspace*{-3mm}\\
$53318.558 \pm 0.133$ & $21.63^{+.05}_{-.05}$ & $0.61^{+.02}_{-.02}$ & $(12.04^{+1.49}_{-1.28}) \times 10^{-3}$ & $1.64^{+.16}_{-.12}$ & $(2.93^{+.40}_{-.42}) \times 10^{-3}$ & 1.24 (432) & $2.98 \times 10^{-12}$ & $1.60 \times 10^{-11}$ \\
\vspace*{-3mm}\\
$53328.543 \pm 0.145$ & $21.56^{+.04}_{-.05}$ & $0.63^{+.02}_{-.01}$ & $(10.89^{+0.96}_{-0.96}) \times 10^{-3}$ & $1.98^{+.19}_{-.15}$ & $(2.19^{+.26}_{-.27}) \times 10^{-3}$ & 1.06 (580) & $2.84 \times 10^{-12}$ & $1.58 \times 10^{-11}$ \\
\vspace*{-3mm}\\
$53338.506 \pm 0.133$ & $21.53^{+.07}_{-.08}$ & $0.62^{+.02}_{-.02}$ & $(9.68^{+1.48}_{-1.27}) \times 10^{-3}$ & $1.59^{+.15}_{-.12}$ & $(3.23^{+.41}_{-.44}) \times 10^{-3}$ & 1.19 (384) & $2.91 \times 10^{-12}$ & $1.60 \times 10^{-11}$ \\
\vspace*{-3mm}\\
$54220.366 \pm 0.167$ & $21.79^{+.04}_{-.04}$ & $0.65^{+.03}_{-.02}$ & $(16.23^{+2.41}_{-2.11}) \times 10^{-3}$ & $1.76^{+.24}_{-.14}$ & $(4.99^{+.71}_{-.80}) \times 10^{-3}$ & 0.91 (458) & $4.37 \times 10^{-12}$ & $1.69 \times 10^{-11}$ \\
\vspace*{-3mm}\\
$54224.180 \pm 0.173$ & $21.76^{+.04}_{-.04}$ & $0.63^{+.02}_{-.02}$ & $(19.99^{+2.88}_{-2.26}) \times 10^{-3}$ & $1.79^{+.22}_{-.11}$ & $(6.39^{+.68}_{-.90}) \times 10^{-3}$ & 1.06 (614) & $5.51 \times 10^{-12}$ & $2.20 \times 10^{-11}$ \\
\vspace*{-3mm}\\
\hline
\end{tabular}
\end{center}
\end{table*}
We tested single temperature fits, but these failed to reproduce the hard part of the spectrum. Significantly better results were achieved for two-temperature fits (see Fig.\,\ref{fit}). The lower temperature is found consistently to be around $0.63 \pm 0.01$\,keV, whilst the hotter component at kT$_2 = (1.74 \pm 0.14)$\,keV is somewhat less constrained. The observed flux varies by more than a factor of two. This is unlikely to be due to a modulation of the circumstellar absorption, since (1) N$_{\rm wind}$ varies relatively little and (2) N$_{\rm wind}$ is at its second highest value when the observed flux reaches its maximum. According to our fits, the flux variations are most likely due to a changing emission measure of the X-ray emitting plasma.

We compared the X-ray fluxes of the various pointings over seven narrow energy bands (0.25 -- 0.6, 0.6 -- 1.0, 1.0 -- 1.5, 1.5 -- 2.0, 2.0 -- 3.0, 3.0 -- 5.0, and 5.0 -- 10.0\,keV). Whilst the four {\it XMM-Newton} spectra taken in 2004 do not show variations of more than 10\% in any of the energy bands, the fluxes appear much higher in the two 2007 spectra, in some bands by more than a factor of 2. Comparing the last two observations with the first four, the increase is by far the most important at energies above 1\,keV and increases with energy. This underlines that the spectrum in our 2007 pointings is significantly harder than those in 2004 as can already be seen by the increase in the norm of the hotter component in the two-temperature fits in Table\,\ref{fitX}. 

To answer the question of whether or not the observed variability is related to the orbital period of the binary system, we also analysed archival {\it ROSAT} and {\it ASCA} spectra.  

Previous studies of these data yielded conflicting conclusions about the X-ray variability of Cyg\,OB2 \#5. Comparison between the {\it ROSAT} observations and {\it EINSTEIN} data in fact suggested that the star had apparently shown two events of enhanced (increase by a factor 2 in flux) X-ray emission between late 1979 and early 1980 (see Waldron et al.\ \cite{Waldron}) as well as in the mid 1990s ({\it ROSAT} All Sky Survey data). Waldron et al.\ (\cite{Waldron}) pointed out that since the star was in its low level radio state during the first event and in a high level state during the second event, these X-ray events are unlikely to be related to the radio emission. Finally, Corcoran (\cite{Mike}) analysed the phase-dependence of the soft X-ray count rate from pointed {\it ROSAT}-PSPC observations. No significant modulation was found for Cyg\,OB2 \#5. 

We thus analysed the {\it ROSAT}-PSPC and {\it ASCA}-SIS spectra, using a single temperature model for the former and a two-temperature model for the latter\footnote{Given the limited energy range of the PSPC instrument, a second temperature would not be constrained by the data and is actually not required to fit the spectrum. However, the hard part of the SIS spectrum can only be fitted by including a second temperature, although the latter is poorly constrained due to the limited quality of the SIS data.}. These fits were then used to evaluate the observed fluxes in the 0.5 -- 2.0\,keV domain. The results are listed in Table\,\ref{Xflux}. The first and second columns yield the dates of the observations (at mid-exposure and with the uncertainties corresponding to half the integration time) and the corresponding orbital phases computed from our quadratic ephemeris. The third column indicates the instrument used in the observation and the last column provides the observed flux in the 0.5 -- 2.0\,keV energy range. 

As far as the {\it ROSAT}-PSPC data are concerned, we find indeed surprisingly little variability: the flux remains constant within 5\% of its mean level. At orbital phases that were observed with all three instruments, the agreement between PSPC, SIS, and EPIC fluxes was extremely good given the very different sensitivities of these devices and the fact that the {\it ASCA}-SIS observation is limited by stray-light contamination from Cyg\,X-3\footnote{We emphasize that, to within 5\%, we obtain the same flux as found by Kitamoto \& Mukai (\cite{KM}) in their analysis of the {\it ASCA}-SIS CCD spectra.}.

\begin{table}[htb]
\caption{The soft (0.5 -- 2.0\,keV) X-ray flux of Cyg\,OB2 \#5 as a function of time. \label{Xflux}}
\begin{center}
\begin{tabular}{c c c c}
\hline
JD - 2400000 & $\phi$ & Instrument & f$_X^{\rm obs}$ \\
             &      &            & erg\,cm$^{-2}$\,s$^{-1}$ \\
\hline
\vspace*{-3mm}\\
$48368.064 \pm 0.137$ & $0.570 \pm 0.021$ & PSPC         & $1.59 \times 10^{-12}$ \\
\vspace*{-3mm}\\
$49106.963 \pm 0.466$ & $0.558 \pm 0.071$ & SIS0 \& SIS1 & $1.66 \times 10^{-12}$ \\
\vspace*{-3mm}\\
$49107.121 \pm 0.087$ & $0.582 \pm 0.013$ & PSPC         & $1.48 \times 10^{-12}$ \\
\vspace*{-3mm}\\
$49109.234 \pm 0.058$ & $0.902 \pm 0.009$ & PSPC         & $1.63 \times 10^{-12}$ \\
\vspace*{-3mm}\\
$49110.131 \pm 0.203$ & $0.038 \pm 0.031$ & PSPC         & $1.62 \times 10^{-12}$ \\
\vspace*{-3mm}\\
$49111.230 \pm 0.376$ & $0.205 \pm 0.057$ & PSPC         & $1.64 \times 10^{-12}$ \\
\vspace*{-3mm}\\
$53308.579 \pm 0.122$ & $0.354 \pm 0.018$ & EPIC         & $1.39 \times 10^{-12}$ \\
\vspace*{-3mm}\\
$53318.558 \pm 0.133$ & $0.866 \pm 0.020$ & EPIC         & $1.57 \times 10^{-12}$ \\
\vspace*{-3mm}\\
$53328.543 \pm 0.145$ & $0.380 \pm 0.022$ & EPIC         & $1.46 \times 10^{-12}$ \\
\vspace*{-3mm}\\
$53338.506 \pm 0.133$ & $0.890 \pm 0.020$ & EPIC         & $1.51 \times 10^{-12}$ \\
\vspace*{-3mm}\\
$54220.366 \pm 0.167$ & $0.543 \pm 0.025$ & EPIC         & $1.98 \times 10^{-12}$ \\
\vspace*{-3mm}\\
$54224.180 \pm 0.173$ & $0.121 \pm 0.026$ & EPIC         & $2.44 \times 10^{-12}$ \\
\vspace*{-3mm}\\
\hline
\end{tabular}
\end{center}
\end{table}
\begin{figure}[htb!]
\resizebox{9.0cm}{!}{\includegraphics{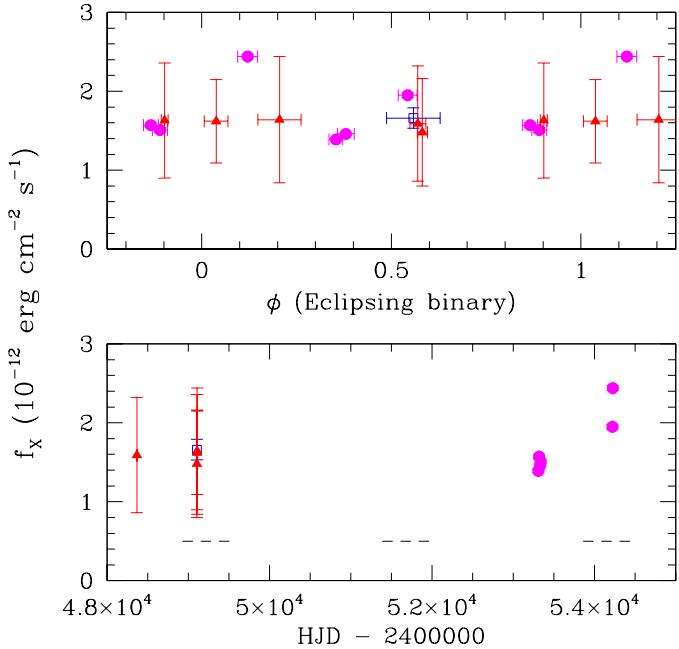}}
\caption{The soft band (0.5 -- 2.0\,keV) flux of Cyg\,OB2 \#5 as a function of orbital phase of the eclipsing binary (top panel) and as a function of time (bottom panel). The filled circles indicate our {\it XMM-Newton} EPIC data, the triangles stand for the {\it ROSAT}-PSPC data whilst the open square corresponds to the {\it ASCA}-SIS observation. The dashed lines indicate the epochs of the minimum radio flux at 6\,cm according to Kennedy et al.\ (\cite{Kennedy}).\label{historicalX}}
\end{figure}

From Fig.\,\ref{historicalX}, we note that the {\it XMM-Newton} observation with the highest X-ray flux was taken when the secondary star was moving away from us. However, it seems rather unlikely that we are dealing with a phase-locked variability since the {\it EINSTEIN}-IPC observation in which the system was detected at a similarly high flux level (Waldron et al.\ \cite{Waldron}) was spread over five days whilst the {\it ROSAT} All Sky Survey data (when the star was also at a high X-ray level, Waldron et al.\ \cite{Waldron}) were collected over an even longer time interval. Therefore, we conclude that there is currently no evidence for a phase-locked modulation of the X-ray flux, at least not with the period of the eclipsing binary system. 
 
An interesting result is the lack of eclipses in the X-ray domain. Indeed, our observations reveal no change in the observed X-ray flux near the phases of the eclipses (Fig.\,\ref{historicalX}). Therefore, the X-ray emitting region must be far wider than the typical dimensions of the stars in the eclipsing binary. A similar situation was reported for $\beta$\,Lyrae (B6-8\,IIp + B0\,V, Ignace et al.\ \cite{betaLyr}), for which it was suggested that the X-rays arise in the wind of the B0 mass-gainer star and that the constancy of the X-ray flux is due to scattering in an extended halo from above and below the disk around the B0 star (Ignace et al.\ \cite{betaLyr}). In the case of Cyg\,OB2 \#5, the X-ray emission could originate in various places. First, it could be intrinsic to the outer regions (several stellar radii) of the primary star's atmosphere. However, the clear detection of a rather hard component in the spectrum argues against this explanation. It could also, at least partially, originate from the wind-wind interaction between the two stars. Since the stars are in a contact configuration, we emphasize that we would expect this X-ray emission to be rather soft if anything, because the winds would interact at velocities far lower than the terminal velocity. We would therefore expect at least some modulation of the wind column density between phases 0.0 (secondary wind in front) and 0.5 (primary wind in front). The existence of a hard component and the lack of significant modulation of the wind column density thus argue against this scenario. The most likely interpretation is then that at least the hard part of the X-ray emission originates in the interaction region between the combined stellar winds of the eclipsing binary and the wind of the tertiary component or the wind of a yet unknown fourth component, responsible for the 6.7\,yr modulation of the radio emission (Kennedy et al.\ \cite{Kennedy}). As far as the former possibility is concerned, it remains to be verified whether the emission measure of this interaction region in such a wide system with a period of centuries would be sufficient to produce a significant hard component in the spectrum. Concerning the latter possibility, the plot of the soft X-ray flux as a function of time in Fig.\,\ref{historicalX} shows that during the 6\,cm radio minimum state (depicted by the dashed lines in Fig.\,\ref{historicalX}), we observe both, low and high X-ray fluxes. Therefore, it seems unlikely a priori that the X-ray variations have the same origin and timescale as the radio variations. We also note the striking similarities between the X-ray spectra of Cyg\,OB2 \#5 and the triple system HD\,167971 that consists of a close eclipsing binary with a third component with a period of several years (see De Becker et al.\ \cite{hd167971}). 

If we adopt a distance of 925\,pc, the average (ISM-corrected) X-ray luminosity in the 0.4 -- 10.0\,keV band would be $1.73 \times 10^{33}$\,erg\,s$^{-1}$. Compared to our estimate of the bolometric luminosity of the eclipsing binary, we thus obtain $\log{\frac{L_{\rm X}}{L_{\rm bol}}} = -6.39$. The system thus appears slightly (by a factor of 3) X-ray overluminous compared with the $L_{\rm X} - L_{\rm bol}$ relation established by Sana et al.\ (\cite{Sana}) for the O-type stars in NGC\,6231. Whilst this is only a mild overluminosity, it is consistent with our interpretation that part of the X-rays originate from wind - wind interactions.

Finally, we have also used the {\it XMM-Newton} data to search for short-term X-ray variability (on the timescale of individual pointings). Hydrodynamical instabilities in the plasma in a colliding-wind region could in principle create these short timescale variations (Stevens et al.\ \cite{SBP}). We considered the EPIC light curves with binnings of 100, 500, 1000, and 4000\,s, in three different energy bands: [250:1000]\,eV, [1000:2500]\,eV, and [2500:10000]\,eV. We applied three independent statistical tests to the data ($\chi^2$, Kolmogorov-Smirnov and probability of variability test; see Sana et al.\ \cite{HD152248}) to detect variations on the timescale of one observation. No significant variability could be detected in any of the energy bands, for any instrument, and/or any time binning.
 
\section{Discussion \label{discussion}}
\subsection{Evidence for a changing period?}
We begin by considering the meaning of the non-linear ephemeris derived from the time of minima. First of all, we emphasize that a linear ephemeris definitely does not provide a good fit to the times of primary minimum. However, since there are only a limited number of extensive (i.e.\ covering several cycles) observing campaigns of this system, the quadratic ephemeris clearly requires confirmation. This is especially true since V\,382\,Cyg, another eclipsing O + O overcontact binary, apparently displays some stochastic variations in its period (Harries et al.\ \cite{Harries}, and references therein) and we cannot exclude the existence of a similar phenomenon in the case of our target. This being said, we have found that the time derivative of the orbital period of Cyg\,OB2 \#5 amounts to $\dot{P} = (18.5 \pm 5.0) \times 10^{-9}$\,s\,s$^{-1} = (0.58 \pm 0.16)$\,s\,yr$^{-1}$. Although this value appears rather large, it is comparable to those values inferred for other evolved eclipsing binaries such as V\,444\,Cyg (WN5 + O6, $\dot{P} = 0.12$\,s\,yr$^{-1}$; Eaton \& Henry \cite{EH}), GP\,Cep (WN6/WCE + O3-6, $\dot{P} = 1.3$\,s\,yr$^{-1}$; Demers et al.\ \cite{Demers}) or CQ\,Cep (WN7 + O7, $\dot{P} = -0.014$\,s\,yr$^{-1}$; Antokhina et al.\ \cite{Antokhina}). 
 
Wellstein (\cite{Wellstein}) pointed out that the detection of changes in the orbital period of Cyg\,OB2 \#5 could help constrain the mass-loss rate and hence the evolutionary stage of this system. However, the interpretation of the rate of period change depends strongly on the way that mass is exchanged inside or lost from the system. Depending on the details of the process, the orbital period can either increase or decrease as a function of time (Khaliullin \cite{Khaliullin}; Singh \& Chaubey \cite{SC}). If we assume that mass is transferred from the primary towards the secondary without any change in the orbital angular momentum and the systemic mass (case I of Singh \& Chaubey \cite{SC}), we find that 
$$\dot{m}_p = \frac{1}{3}\,\frac{m_p\,m_s}{(m_p - m_s)}\,\frac{\dot{P}}{P}$$
This then yields a mass-transfer rate of $\dot{m}_p = +(4.7 \pm 1.5) \times 10^{-6}$\,M$_{\odot}$\,yr$^{-1}$, i.e.\ mass would be transferred from the secondary towards the primary star. As we argue below, this seems rather unlikely.

Alternatively, we can assume that the mass is lost primarily through an isotropic (to first order) stellar wind (case IV of Singh \& Chaubey \cite{SC}). In this situation, the mass-loss rate can be directly computed to be
$$\dot{m}_p = -\frac{1}{2}\,(m_p + m_s)\,\frac{\dot{P}}{P}$$
In our case, we thus obtain a mass-loss rate from the binary system of $\dot{m}_p = -(2.1 \pm 0.6) \times 10^{-5}$\,M$_{\odot}$\,yr$^{-1}$ quite similar to the value inferred for the Wolf-Rayet component of GP\,Cep ($\dot{M} = -2.9\,10^{-5}$\,M$_{\odot}$\,yr$^{-1}$; Demers et al.\ \cite{Demers}) and about a factor of 3 higher than the most extreme values found for V\,444\,Cyg (Eaton \& Henry \cite{EH}).

We emphasize that in the particular case of Cyg\,OB2 \#5 an additional effect could alter the observed times of minimum light. This effect is related to the third component dicussed by Contreras et al.\ (\cite{Contreras}). As described in Sect.\,\ref{introduction}, these authors confirmed the presence of a third star, probably of spectral type B0-2\,V, at an angular distance of 0.98\,arcsec from the close binary. If this third star is gravitationally bound to the binary system and if the angular separation corresponds to the radius of a circular orbit or the semi-major axis of an elliptical orbit, the corresponding orbital period would be about 9200\,yr for a distance of 1.7\,kpc (or 3700\,yr for a distance of 925\,pc). The separation between the epoch of Hall (\cite{Hall}) and our data is about 29.1\,yr. If we make yet another assumption that the triple system is seen under an inclination of $90^{\circ}$ and is currently close to quadrature phase, then the maximum change in the distance of the eclipsing binary towards the Earth would be about $5 \times 10^{9}$\,km (or about $7 \times 10^{9}$\,km if the system lies at 925\,pc). Due to the finite velocity of light, such a change in distance can produce a lag in the time of minimum light. Interpreting the $0.266$\,day delay between Hall's linear ephemeris and our determination of the time of minimum in this way, requires a change in distance of $6.9 \times 10^{9}$\,km. Whilst there are admittingly a large number of assumptions here, we nevertheless note that this light-travel effect could contribute significantly to the observed deviation from the old linear ephemeris. 

In summary, it is probable that several effects (mass-transfer, mass-loss, and orbital motion in a triple system) conspire to produce a non-linear ephemeris in Cyg\,OB2 \#5. Additional accurate determinations of the primary minimum over a long timescale would be most welcome in clarifying this issue.

\subsection{The interactions in the eclipsing binary system}
An important result of our study is that the components of the eclipsing binary are currently in a contact configuration. In addition, we find that the secondary appears hotter over part of its surface and this is probably due to the energy transfer in a common envelope and the heating of the secondary star by both the primary's radiation field and the impact of the primary's stellar wind on the secondary's surface. It is interesting to compare this configuration with the radiation-hydrodynamics simulations of Dessart et al.\ (\cite{Dessart}). These authors simulated mass transfer in a binary system in the presence of strong stellar winds and showed that neither the stellar wind's momentum nor the radiation field of the accreting star can prevent mass transfer by Roche lobe overflow. Dessart et al.\ (\cite{Dessart}) further showed that the wind of the accreting star is always dominated by the primary's wind and the mass stream. The fact that the Doppler maps of Cyg\,OB2 \#5 and the S-wave analyses performed by Rauw et al.\ (\cite{Rauw}) indicated that (line-emitting) material is moving towards the secondary hence suggests that the primary is transferring matter to its companion. 

Our finding that the X-ray flux of the system is not modulated by the short period of the eclipsing binary indicates that the bulk of the X-ray emissions does not originate from the wind-wind interaction between the two components. It rather seems that the interaction between the wind of the eclipsing system and that of a tertiary or maybe even a fourth component could be at the origin of the X-ray emission. It is likely that this also holds for the acceleration of relativistic particles responsible for the non-thermal radio emission (Kennedy et al.\ \cite{Kennedy}) and the potential contribution to the yet unidentified {\it EGRET} $\gamma$-ray (100\,MeV -- 20\,GeV) source 3EG\,J2033+4118 (Benaglia et al.\ \cite{Benaglia})\footnote{Concerning the latter, we note that {\it INTEGRAL} observations (De Becker et al.\ \cite{DeBecker}) failed to detect a hard X-ray, low $\gamma$-ray counterpart to the {\it EGRET} source.}.

\subsection{The evolutionary state}
An important consequence of our finding that the system is currently in a contact configuration (as originally suggested by Leung \& Schneider \cite{LS}), is the resolution of the apparent conflict between the luminosity ratio and the mass ratio. Our photometric solution indeed yields a luminosity ratio close to 3.1. This is well below the value (by a factor of 100 or more!) expected from the dynamical mass ratio of 3.3, if we were dealing with single stars or components of an unevolved detached binary system! We note also that the secondary's absorption lines yield an optical brightness ratio of $1.4 \pm 0.6$ (Rauw et al.\ \cite{Rauw}) which is in marginal agreement with the value of 2.9 inferred from the photometric solution. 

The key to this puzzle probably lies in the fact that the two stars are in a contact configuration. In their pioneering work to explain the unusual mass-luminosity relations of W\,UMa-type contact binaries, Osaki (\cite{Osaki}) and Lucy (\cite{Lucy}) showed that the fact that energy is radiated according to a gravity darkening law implies that in a common envelope system $\frac{L_1}{L_2} \simeq \frac{M_1}{M_2}$, which agrees with our results for Cyg\,OB2 \#5. These authors also showed that both stars in a contact binary system should have almost equal surface brightnesses, which is indeed what we observe in Cyg\,OB2 \#5. Whilst Osaki (\cite{Osaki}) adopted the conventional von Zeipel gravity darkening law that applies to stars with a radiative envelope, Lucy (\cite{Lucy}) argued that the von Zeipel law is inappropriate for W\,UMa contact binaries that are typically of spectral type K -- F and should thus have a convective common envelope. Lucy (\cite{Lucy}) also questionned the possibility to apply Osaki's (\cite{Osaki}) conclusions to early-type contact binaries (such as the one we are dealing with in this work), because the common envelope would not be in hydrostatic equilibrium. However, Tassoul (\cite{Tassoul}) showed that early-type contact binaries can have a barotropic outer common envelope hence validating the use of von Zeipel's gravity darkening law. Whilst Cyg\,OB2 \#5 is certainly far hotter and more massive than classical W\,UMa-type systems, it seems likely also in our case that the spectra of both components originate in a common photosphere and this explanation then accounts for many observed properties of the system. However, there is one caveat: our analysis has revealed a secondary's star surface that is apparently not in equilibrium since we found a bright region facing the primary star. Therefore, more sophisticated theoretical models are needed to understand the structure of a common envelope during mass exchange in early-type binaries.

In previous papers, it was argued that the secondary of Cyg\,OB2 \#5 might be a Wolf-Rayet or Ofpe/WNL transition object with a spectrum altered by a hydrogen cloaking due to mass transfer from the primary (Bohannan \& Conti \cite{BC}; Vreux \cite{JMV}; Rauw et al.\ \cite{Rauw}). In the light of the results obtained in this paper, we emphasize that there is no real need for such a scenario anymore. The contact configuration indeed probably accounts for most of the odd features of this star, which could otherwise be a rather `normal' B1-2 star. 
\subsection{The distance of Cyg\,OB2}
Our analysis of the light curve of Cyg\,OB2 \#5 yields a distance of 900 -- 950\,pc, roughly half of the previous distance estimates of Cyg\,OB2. While this result needs to be confirmed, it is interesting to note that this revision of the distance would have important implications for our general understanding of the stars in Cyg\,OB2. First of all, the luminosities (and luminosity classes) of the O stars would have to be revised. A reduction in the luminosities and thus also the mass-loss rates of the stars would then have a potential impact on our understanding of the radio light-curve of the non-thermal radio emitting O-stars such as Cyg\,OB2 \#8a and \#9 (De Becker et al.\ \cite{Michael}; Naz\'e et al.\ \cite{Yael}). In these binary systems with relatively short orbital periods, the wind interaction zone is rather deeply embedded in the winds of the stars. With current estimates of the mass-loss rates, the bulk of the synchrotron radio emission produced in this wind collision region would be absorbed by the winds themselves. A downward revision of the distance of these stars and their wind densities would imply a lower opacity in the radio domain and could help to solve this issue. 
\section{Conclusion \label{conclusion}}
Using optical narrow-band photometry and spectroscopy, we have shown that Cyg\,OB2 \#5 is in a contact configuration, which has allowed us to solve one of the oldest mysteries associated with this system, i.e.\ the apparent discrepancy between its mass ratio and its luminosity ratio. The binary appears to be in a rather `normal' stage of (likely case-A) mass transfer in binary evolution. 

However, whilst we have made progress in understanding the system, there remain a number of major issues. The most important ones are:
\begin{itemize}
\item[$\bullet$] New theoretical models need to be designed to simulate accurately the structure of a common envelope in an early-type, contact binary system. 
\item[$\bullet$] The origin of the period variations needs to be investigated by long-term photometric monitoring of the system. 
\item[$\bullet$] The origin of the X-ray emission in this system remains uncertain as is the cause of the variations in the radio emission (Kennedy et al.\ \cite{Kennedy}). The most likely explanations involve the existence of a fourth component in the system. However, to date no direct evidence that such a component exists has been obtained and again long term monitoring of the system is required. For instance, one needs to acquire optical spectroscopy to determine orbital solutions that sample the 6.7\,yr cycle.  
\item[$\bullet$] And finally, the distance of Cyg\,OB2 needs to be checked by studying more eclipsing binary systems in this important cluster. 
\end{itemize}
\acknowledgement{We are grateful to the referee, Dr.\ A.F.J.\ Moffat for helpful comments, to Dr.\ H.\ Sana who took part in the 1998 OHP photometric observing campaign as well as to Dr.\ S.\ Dougherty and M.\ Kennedy for sharing their results on the radio data of Cyg\,OB2 \#5 with us before publication. The Li\`ege group acknowledges financial support from the FRS/FNRS (Belgium), as well as through the XMM and INTEGRAL PRODEX contract (Belspo). The travels to OHP were supported by the `Communaut\'e Fran\c caise' (Belgium). PE acknowledges support through CONACyT grant 67041.}

\end{document}